\documentclass[a4paper]{article}
\usepackage{authblk}

\advance\textheight by 1cm
\advance\topmargin -1cm

\advance\textwidth by 2cm
\advance\oddsidemargin by -1cm
\advance\evensidemargin by -1cm

\usepackage[utf8]{inputenc}
\usepackage[T1]{fontenc}

\usepackage{pdfsync}
\usepackage{amsmath, amssymb}
\usepackage{graphicx} 
\usepackage[thmmarks]{ntheorem}
\usepackage[hyphens]{url}
\usepackage{bm}
\usepackage{todonotes}
\usepackage{booktabs}

\def\..{\,\mathpunct{\ldotp\ldotp}} 

\usepackage{microtype} 
\usepackage{algorithm}
\usepackage[noend]{algorithmic}
\usepackage{authblk}

\usepackage{color}
\usepackage{listings}
\lstdefinestyle{customc}{
  belowcaptionskip=1\baselineskip,
  breaklines=true,
  xleftmargin=\parindent,
  language=C++,
  showstringspaces=false,
  basicstyle=\small\ttfamily,
  commentstyle=\color[rgb]{0.3, 0.3, 0.4}
}

\newcommand{\F}{\mathbf F}

\newcommand{\ttlpar}[1]{\noindent{\bfseries\boldmath #1}\enspace}

\newcommand{\GF}[1]{\mathbf F_#1}

\date{}
\author[1]{Marco Genuzio}
\author[2]{Giuseppe Ottaviano}
\author[1]{Sebastiano Vigna\thanks{Sebastiano Vigna and Marco Genuzio are
supported by a Google Focused Grant.}} \affil[1]{Dipartimento di Informatica, Universit\`a degli Studi di Milano,
Milan, Italy}
\affil[2]{Facebook, Menlo Park, USA}
\title{Fast Scalable Construction of (Minimal Perfect Hash) Functions}

\begin{document}
\bibliographystyle{plain}
\maketitle
\begin{abstract}
  Recent advances in random linear systems on finite fields have paved the way
  for the construction of constant-time data structures representing static
  functions and minimal perfect hash functions using less space with respect to
  existing techniques. The main obstruction for any practical application of
  these results is the cubic-time Gaussian elimination required to solve these
  linear systems: despite they can be made very small, the computation is still
  too slow to be feasible.

  In this paper we describe in detail a number of heuristics and programming
  techniques to speed up the resolution of these systems by several orders of
  magnitude, making the overall construction competitive with the standard and
  widely used MWHC technique, which is based on hypergraph peeling. In
  particular, we introduce \emph{broadword programming} techniques for fast
  equation manipulation and a \emph{lazy Gaussian elimination} algorithm. We
  also describe a number of technical improvements to the data structure which
  further reduce space usage and improve lookup speed.
  
  Our implementation of these techniques yields a minimal perfect hash function
  data structure occupying $2.24$ bits per element, compared to $2.68$ for
  MWHC-based ones, and a static function data structure which reduces the
  multiplicative overhead from $1.23$ to $1.03$.
\end{abstract}

\section{Introduction}\label{sec:intro}
\emph{Static functions} are data structures designed to store arbitrary
mappings from finite sets to integers; that is, given a set of $n$ pairs $(k_i,
v_i)$ where $k_i \in S \subseteq U, |S|=n$ and $v_i \in 2^b$, a static
function will retrieve $v_i$ given $k_i$ in constant time. Closely related are
\emph{minimal perfect hash functions (MPHFs)}, where only the set $S$ of $k_i$'s
is given, and the data structure produces an injective numbering $S\to n$. 
While these tasks can be easily implemented using hash tables,
static functions and MPHFs are allowed to return \emph{any} value if the 
queried key is not in the original set $S$; this relaxation enables to break the
information-theoretical lower bound of storing the set $S$. In fact,
constructions for static functions achieve just $O(nb)$ bits space and MPHFs
$O(n)$ bits space, regardless of the size of the keys. This makes static
functions and MPHFs powerful techniques when handling, for instance, large sets
of strings, and they are important building blocks of space-efficient data
structures such as (compressed) full-text indexes~\cite{BeNACTI}, monotone
MPHFs~\cite{BBPMMPH,BBPTPMMPH2}, Bloom filter-like data structures~\cite{BeVCSFA}, and
prefix-search data structures~\cite{BBPFPSLSA}.

An important line of research, both theoretical and practical, involves lowering
the multiplicative constants in the big-$O$ space bounds, while keeping feasible
construction times. In this paper we build on recent advances in random linear
systems theory, and in perfect hash data structures~\cite{DiPSDSRAM,RinPHD}, to
achieve practical static functions with the lowest space bounds so far, and
construction time comparable with widely used techniques. The new results,
however, require solving linear systems rather than a simple depth-first visit
of a hypergraph, as it happens in current state-of-the-art solutions.

Since we aim at structures that can manage billions of keys, the main challenge
in making such structures usable is taming the cubic running time of Gaussian
elimination at construction time. To this purpose, we introduce novel techniques based on \emph{broadword
programming}~\cite{KnuACPBTT} and a lazy version of \emph{structured Gaussian
elimination}. We obtain data structures that are significantly smaller than
widely used hypergraph-based constructions, while maintaining or improving the lookup
times and providing still feasible construction time.

All implementations discussed in this paper are distributed as free software as
part of the Sux4J project (\texttt{http://sux4j.di.unimi.it/}).

\section{Notation and tools}
\label{sec:notation}

We use von Neumann's definition and notation for natural numbers,
identifying $n$ with $\{\,0,1,\ldots,n-1\,\}$, so $2=\{\,0,1\,\}$ and
$2^b$ is the set of $b$-bit numbers.

\ttlpar{Model and assumptions}
Our model of computation is a unit-cost word RAM with word size
$w$. We assume that $n=|S|=O(2^{cw})$ for some constant $c$, so that
constant-time static data structures depending on $|S|$ can be used.

\ttlpar{Hypergraphs}%
An $r$-hypergraph on a vertex set $V$ is a subset $E$ of ${V \choose r}$, the
set of subsets of $V$ of cardinality $r$. An element of $E$ is called
an \emph{edge}. The $k$-core of a hypergraph is its maximal induced subgraph
having degree at least $k$.

A hypergraph is \emph{peelable} if it is possible to sort its edges in a list so
that for each edge there is a vertex that does not appear in following elements
of the list. A hypergraph is peelable if and only if it has an empty $2$-core.
It is \emph{orientable} if it is possible to associate with each hyperedge a
distinct vertex. Clearly, a peelable hypergraph is orientable, but the converse
is not necessarily true.

\section{Background and related work}
\ttlpar{Linear functions and MWHC.} Most static function constructions work by
finding a \emph{linear function} that satisfies the requirements. For simplicity start with the special case of functions with binary values, that is $v_i \in \F_2$ (the field with two elements); the task is to find a vector $w \in \F_2^m$ such that for each $i$
\begin{equation}
  h_\theta(k_i)^T w = v_i
\end{equation}
where $h_\theta$ is a function $U
\rightarrow \F_2^m$ from a suitable family $\mathcal{H}$ indexed by $\theta$. 
To make the lookup constant-time, we add the
additional constraint that $h_\theta(k)$ has a constant number $r$ of 
ones, and that the positions of these ones can be
computed in constant time. Then, with a slight
abuse of notation, we can write $h_{\theta, j}$ to be the \emph{position} of the
$j$-th nonzero element, and hence the lookup just becomes
\begin{equation}
 w_{h_{\theta,
0}(k_i)} + \dots + w_{h_{\theta, r-1}(k_i)} = v_i.
\end{equation}

It is clear that, if such a function exists, the data structure just requires to
store $w$ and $\theta$. Note that if $h_\theta$ is fixed, just writing down the
$n$ equations above yields a linear system: stacking the row vectors
$h_\theta(k_i)^T$ into a matrix $H$ and the values $v_i$ into the vector $v$, we
are looking to solve the equation
\begin{equation}
  H w = v \text{.}
\end{equation}
A sufficient condition for the solution $w$ to exist is that the matrix $H$ has
full rank. To generalize to the case where $v_i \in \F_2^b$ is a $b$-bit
integer, just replace $v$ with the $n \times b$ matrix $V$ obtained by stacking
the $v_i$'s as rows, and $w$ by a $m \times b$ matrix. Full rank of $H$ is
still a sufficient condition for the solvability of $HW = V$. It remains to show
how to pick the number of variables $m$, and the functions $h_\theta$, so that
$H$ has full rank.

In their seminal paper~\cite{MWHFPHM}, Majewski, Wormald, Havas and Czech (MWHC
hereinafter) introduced the first static function construction that can be
described using the framework above. They pick as $\mathcal{H}$ the set of
functions $U\to \F_2^m$ whose values have exactly $r$ ones, that is,
$h_\theta(k)$ is the vector with $r$ ones in positions $h_{\theta,j}(k)$ for
$j\in r$, using the same notation above. If the functions are picked uniformly
at random, the $r$-uples $\bigl(h_{\theta, 0}(k), \dots, h_{\theta,
r-1}(k)\bigr)$ can be seen as edges of a random hypergraph with $m$ nodes.
When $m > c_r n$ for a suitable $c_r$, with high probability the hypergraph
is peelable, and the peeling process
\emph{triangulates} the associated linear system; in
other words, we have both a probabilistic guarantee that the system is
solvable, and that the solution can be found in linear time. The constant $c_r$
depends on the degree $r$, which attains its minimum at $r=3$, $c_3 \approx
1.23$. The family $\mathcal{H}$ can be substituted with a smaller set where the
parameter $\theta$ can be represented with a sublinear number of bits, so the
overall space is $1.23 b n + o(n)$ bits. In practice, $h_{\theta, j}(k)$ will be
simply a hash function with random seed $\theta$, which can be represented in
$O(1)$ bits.

\ttlpar{MPHFs.} Chazelle, Kilian, Rubinfeld
and Tal~\cite{CKRBF}, unaware of the MWHC construction,
proposed it independently, but also noted that as a side-effect of the peeling
process each hyperedge can be assigned an unique node; that is, each key can be
assigned injectively an integer in $m$. We just need to store which of the $r$
nodes of the hyperedge is the assigned one  to obtain a perfect hash
function $S \to m$, and this can be done 
in $c_r\lceil \log r \rceil n + o(n)$ bits. To make it perfect, that is, $S \to
n$, it is possible to add a ranking structure. Again, the best $r$ is $3$, which yields theoretically 
a $2.46n + o(n)$ data structure~\cite{BPZPPHNOS}.

\ttlpar{HEM.} Botelho, Pagh and Ziviani~\cite{BPZPPHNOS} introduced a practical
external-memory algorithm called Heuristic External Memory (HEM) to construct MPHFs
for sets that are too large to store their hypergraph in memory. They replace each
key with a \emph{signature} of $\Theta(\log n)$ bits computed with a random hash
function, and check that no collision occurs. The signatures are then
sorted and divided into small chunks based on their most significant bits, and a separate
function is computed for each chunk with the approach described above (using
a local seed). The representations of the chunk functions are then concatenated into a
single array and their offsets (i.e., for each chunk, the position of the start
of the chunk in the global array) are stored separately.

\ttlpar{Cache-oblivious constructions.} As an alternative to HEM, in \cite{BBOCOPRH}
the authors propose \emph{cache-oblivious} algorithms that use only scanning and
sorting to peel hypergraphs and compute the corresponding structures. The main
advantage is that of avoiding the cost of accessing the offset array of HEM without
sacrificing scalability.

\ttlpar{CHD.} Finally, specifically for the purpose of computing MPFHs
Belazzougui, Botelho and Dietzfelbinger~\cite{BBDHDC} introduced a completely
different construction, called CHD (compressed hash-and-displace), which, at the
price of increasing the expected construction time makes it possible, in theory, to reach the
information-theoretical lower bound of $\approx 1.44$ bits per key.

\ttlpar{Beyond hypergraphs.} The MWHC construction for static functions can be
improved: Dietzfelbinger and Pagh~\cite{DiPSDSRAM} introduced a new construction
that allows to make the constant in front of the $nb$ space bound for static
functions \emph{arbitrarily small}; by Calkin's theorem, a constant $\beta_r$
exists such that if $m > \beta_r n$ and the rows of the matrix $H$ are just
drawn at random from vectors of weight $r$ then $H$ has full rank with high
probability. Contrary to $c_r$ which has a finite minimum, $\beta_r$ vanishes
quickly as $r$ increases, thus the denser the rows, the closer $m$ can be to
$n$. For example, if $r=3$, $\beta_3 \approx 1.12 < c_3 \approx 1.23$. Unlike
MWHC's linear-time peeling algorithm, general matrix inversion requires
super\emph{quadratic} time ($O(n^3)$ with Gaussian elimination); to obtain a
linear-time algorithm, they shard the set $S$ into small sets using a hash
function, and compute the static functions on each subset independently; the
actual construction is rather involved, to account for some corner cases (note
that the HEM algorithm described above is essentially a practical simplified
version of this scheme).

The authors also notice that solvability of the system implies that the
corresponding hypergraph is orientable, thus making it possible to construct
minimal perfect hash functions. Later
works~\cite{FountoulakisP12,FriezeM12,DGMMPRTTCHX} further improve the
thresholds for solvability and orientability: less than $1.09$ for $r=3$, and
less than $1.03$ for $r=4$.

\section{Squeezing space}

In this paper, we combine a number of new results and techniques to provide
improved constructions. Our data structure is based on the HEM
construction~\cite{BPZPPHNOS}: the key set is randomly sharded into chunks of
expected constant size, and then the (minimal perfect hash) function is computed
independently on each chunk. Instead of using a vertex/edge ratio that
guarantees peelability, however, we choose a lower one that still guarantees
orientability and solvability of the associated linear system (with high
probability). Losing peelability implies that we have to use Gaussian
elimination to solve the linear system, but since the chunks have constant size
the overall construction is linear-time (plus an $O(n \log n)$ step to sort the
signatures, which is actually a small part of the execution time in practice).
We also describe improvements to the HEM data structure in
Section~\ref{sec:betterhem}.

First of all, we use the orientability thresholds in~\cite{DGMMPRTTCHX}, which
are shown to be the same as those of XORSAT solvability; for example, when a
random $3$-hypergraph has a vertex/edge ratio $c>1.09$, it contains a nonempty
$2$-core (i.e., a maximal subgraph all whose vertices have degree at least $2$),
but the hypergraph is orientable and the incidence matrix has full rank. We can
thus extend the MWHC technique to $3$-hypergraphs with a nonempty $2$-core:
after the peeling procedure, we simply solve the equations specified by the
$2$-core. The main obstacle to this approach, before the results described in
this paper, was that construction time was two orders of magnitude slower than
that of the MWHC construction~\cite{ADREVTGRDS}, making the whole construction
unusable in practice. In Michael Rink's Ph.D.~thesis~\cite{RinPHD} these
considerations are described in some detail.

Moreover, since recently Goerdt and Falke have proved a result analogous to
XORSAT for modulo-3 systems~\cite{GoFSTBkX},\footnote{Technically, the proof in
the paper is for $k>15$, but the author claim that the result can be proved for
$k\geq 3$ with the same techniques, and in practice we never needed
more than two attempts to generate a solvable system.} we can also
obtain an orientation of a random $3$-hypergraph using the
\emph{generalized selfless algorithm}~\cite{DGMMPRTTCHX}, and then 
solve the modulo-3 linear system induced by the orientation to obtain
a perfect hash function. Both procedures have some controlled probability of
failure. In case such a failure occurs, we generate a new
hypergraph.
We then show how to manage the ranking part essentially with no space cost.


\section{Broadword programming for row operations}

Our first step towards a practical solution by Gaussian elimination is
\emph{broadword programming}~\cite{KnuACPBTT} (a.k.a.~SWAR---``SIMD in A
Register''), a set of techniques to process simultaneously multiple values by packing them into
machine words of $w$ bits and performing the computations on the whole words. In
theoretical succinct data structures it is common to assume that
$w = \Theta(\log n)$ and reduce to subproblems of size $O(w)$, whose results can
be precomputed into sublinear-sized tables and looked up in constant time. For
practical values of $n$, however, these tables are far from negligible; in this
case broadword algorithms are usually sufficient to compute the same functions
in constant or near-constant time without having to store a lookup table.

For our problem, the inner loop of the Gaussian elimination is entirely composed of
row operations: given vectors $x$ and $y$, and a scalar $\alpha$, compute $x +
\alpha y$. It is trivial to perform this operation $w$ elements at a time when the
field is $\GF{2}$, which is the case for static functions computation: we can
just pack one element per bit, and since the scalar can be only $1$ the sum is
just a bitwise XOR \lstinline[style=customc]|x ^ y|, using the C notation.
For MPHFs, instead, the field is $\GF{3}$, which requires more sophisticated
algorithms. First, we can encode each element $\{0, 1, 2\}$ into $2$ bits, thus
fitting $w/2$ elements into a word. The scalar $\alpha$ can be only $1$ or $-1$, so we can
treat the cases $x + y$ and $x - y$ separately.

For the addition, we can start by simply adding $x$ and $y$. When elements on both
sides are smaller than $2$, there's nothing to do: the result will be smaller
than $3$. When however at least one of the two is $2$ and the other one is not
$0$, we need to subtract $3$ from the result to bring it back to the canonical
representation in $[0\..3)$. Note that when the two sides are both $2$ the
result overflows its $2$ bits ($10_2 + 10_2 = 100_2$), but since addition and
subtraction modulo $2^w$ are associative we can imagine that the operation is
performed independently on each $2$-bit element, as long as the final result fits
into $2$ bits. Thus we need to compute a mask that is $3$ wherever the results
is at least $3$, and then subtract it from $x + y$.
\begin{lstlisting}[style=customc]
uint64_t add_mod3_step2(uint64_t x, uint64_t y) {
    uint64_t xy = x | y;
    // Set MSB if (x or y == 2) and (x or y == 1).
    uint64_t mask = (xy << 1) & xy;
    // Set MSB if (x == 2) and (y == 2).
    mask |= x & y;
    // The MSB of each 2-bit element is now set 
    // iff the result is >= 3. Clear the LSBs.
    mask &= 0x5555555555555555 << 1;
    // Now turn the elements with MSB set into 3.
    mask |= mask >> 1;
    return x + y - mask;
}
\end{lstlisting}

Subtraction is very similar. We begin by subtracting elementwise $y$ from $3$,
which does not cause any carry since all the elements are strictly smaller than
$3$. The resulting elements are thus at least $1$. We can now proceed to compute
$x + y$ with the same case analysis as before, except now the right-hand
elements are in $[1\..3]$ so the conditions for the mask are slightly different.

\begin{lstlisting}[style=customc]
uint64_t sub_mod3_step2(uint64_t x, uint64_t y) {
    // y = 3 - y.
    y = 0xFFFFFFFFFFFFFFFF - y; 
    // Now y > 0
    // Set MSB if x == 2.
    uint64_t mask = x;
    // Set MSB if (x == 2 and y >= 2) or (y == 3).
    mask |= ((x | y) << 1) & y;
    mask &= 0x5555555555555555 << 1;
    mask |= mask >> 1;
    return x + y - mask;
}
\end{lstlisting}

Both addition and subtraction take just 10 arithmetic operations, and on modern
64-bit CPUs they can process vectors of 32 elements at a time.

Finally, when performing back substitution we will need to compute row-matrix
multiplications, where a row is given by the coefficients of an equation and the
matrix contains the solutions computed so far.

In the field $\GF2$, this can be achieved by iterating on the ones of the row,
and adding up the corresponding $b$-bit rows in the right-hand matrix. The ones
can iterate by finding the LSB of the current row word, and deleting it with
the standard broadword trick \verb|x = x & -x|.

For MPHFs, instead, the field is $\GF3$ but the matrix of solutions is a vector,
so the product is just a scalar product. To compute it, we use the following
broadword algorithm that computes the scalar product of two vectors represented
as 64-bit words.

\begin{lstlisting}[style=customc]
uint64_t prod_mod3_step2(uint64_t x, uint64_t y) {
    uint64_t high = x & 0xAAAAAAAAAAAAAAAA;
    uint64_t low = x & 0x5555555555555555;
    // Make every 10 into a 11 and zero everything else.
    uint64_t high_shift = high >> 1;
    // Exchange ones with twos, and make 00 into 11.
    uint64_t t = (y ^ (high | high_shift)) 
        & (x | high_shift | low << 1);
    return popcount(t & 0xAAAAAAAAAAAAAAAA) * 2
      + popcount(t & 0x5555555555555555);
}
\end{lstlisting}

The expression computing $t$ takes care of placing in a given position a value
equivalent to the product of the associated positions in $x$ and $y$ (this can
be easily check with a case-by-case analysis). We remark that in some cases we
actually use $3$ as equivalent to zero. At that point, the last lines compute
the contribution of each product (\texttt{popcount()} returns the number of bit
in a word that are set). Note that the results has still to be reduced modulo 3.

\section{Lazy Gaussian Elimination}

Even if armed with broadword algorithms, solving by Gaussian elimination systems
of the size of a HEM chunk (thousands of equations and variables) would be
prohibitively slow, making construction of our data structures an order of
magnitude slower than the standard MWHC technique.

\emph{Structured Gaussian elimination} aims at reducing the number of operations 
in the solution of a linear system by trying to isolate a number of variables
appearing in a large number of equations, and then rewrite the rest of the
system using just those variables. It is a heuristics 
developed in the context of computations of discrete logarithms, which require
the solution of large sparse systems~\cite{OdlDLFFCS,LaOSLSLSFF}. The standard
formulation requires the selection of a fraction (chosen arbitrarily) of
variables that appear in a large number of equations, and then a number of
loosely defined refinement steps.

We describe here a new parameterless version of structured Gauss elimination,
which we call \emph{lazy Gaussian elimination}. This heuristics turned out
to be extremely effective on our systems, reducing the size of the system to
be solved by standard elimination to around $4$\% of the original one.

Consider a system of equations on some field. At any time a variable can be
\emph{active}, \emph{idle}, or \emph{solved} and an equation can be
\emph{sparse} or \emph{dense}.
Initially, all equations are sparse and all variables are idle. We will modify
the system maintaining the following invariants:
\begin{itemize}
  \item dense equations do not contain idle variables;
  \item an equation can contain at most one solved variable;
  \item a solved variable appears in exactly one dense equation.
  \end{itemize}
Our purpose is to modify the system so that all equations are dense,
trying to minimize the number of active variables (or, equivalently, maximize
the number of solved variables).
At that point, values for the active variables can be computed by 
standard Gaussian elimination on the dense equations that do not
contain solved variables, and solved variables can be computed easily from the
values assigned to active variables.

The \emph{weight} of a variable is the number of sparse equations in which it
appears. The \emph{priority} of a sparse equation is the number of idle
variables in the equation.
Lazy Gaussian elimination keeps equations in a min-priority queue, and performs
the following actions:
\begin{enumerate}
  \item If there is a sparse equation of priority zero that contains
  some variables, it is made dense. If there are no variables, the
  equation is either an identity, in which case it is discarded, or it is impossible, in which case the
  system is unsolvable and the procedure stops.
  \item\label{step:solved} If there is a sparse equation of priority one, the
  only idle variable in the equation becomes solved, and the equation becomes
  dense. The equation is then used to eliminate the solved variable from all
  other equations.
  \item\label{step:dense} Otherwise, the idle variable appearing in the
  largest number of sparse equations becomes active.
\end{enumerate}
Note that if the system is solvable the procedure always completes---in the
worst case, by making all idle variables active (and thus all equations dense).

Two observations are in order:
\begin{itemize}
  \item The weight of an idle variable never changes, as in
  step~\ref{step:solved} we eliminate the solved variable and modify the
  coefficients of active variables only. This means that we can simply sort
  initially (e.g., by countsort) the variables by the number of equations in 
  which they appear, and pick idle variables in that order at
  step~\ref{step:dense}.
  \item We do not actually need a priority queue for equations: simply, when an
  equation becomes of priority zero or one, it is moved to the left or
  right side, respectively, of a deque that we check in the first step.
\end{itemize}
Thus, the only operations requiring superlinear time are the eliminations
performed in step~\ref{step:solved}, and the final Gaussian elimination on
the dense equations, which we compute, however, using broadword programming.

\section{Data structure improvements}\label{sec:betterhem}

\ttlpar{Improving HEM.} Our HEM version uses \emph{on-disk bucket sorting} to
speed up construction: keys are first divided into 256 on-disk \emph{physical}
chunks, depending on the highest bits of their hash value (we use
Jenkins's SpookyHash).
The on-disk chunks are then loaded in memory and sorted, and virtual chunks of the desired size are computed either
splitting or merging physical chunks. Since we store a 192-bit hash plus a
64-bit value for each key, we can guarantee that the amount of memory used 
that
depends on the number of keys cannot exceed one bit per key (beside the structure to be computed).

\ttlpar{Eliminating the ranking structure.} In the case of minimal perfect
hashing, we can further speed up the structure and reduce space by getting rid
of the ranking structure that is necessary to make minimal the perfect hashing
computed by the system of equations.

In the standard HEM construction, the number of vertices associated to a chunk
of size $s$ is given by $\lceil c s\rceil$, where $c$ is a suitable constant,
and the offset information contains the partial sums of such numbers.

We will use a different approach: the number of vertices associated with the
chunk will be $\lceil c (S+s)\rceil-\lceil c S\rceil$, where $S$ is the number
of elements stored in previous chunks.  The difference to $\lceil c s\rceil$
is at most one, but using our approach we can compute, given $S$ and $s$, the
number of vertices associated with the chunk.

Thus, instead of storing the offset information, we will store for each chunk
the number $S$ of elements stored in previous chunks. The value can be used as a
base for the ranking inside the chunk: this way, the ranking structure is no
longer necessary, reducing space and the number of memory accesses. 
When $r=3$, as it is customary we can use two bits for
each value, taking care of using the value 3, instead of 0, for the vertex
associated to a hyperedge. As a result, ranking requires
just counting the number of nonzero pairs in the values
associated with a chunk, which can be performed again by broadword programming:
\begin{lstlisting}[style=customc]
int count_nonzero_pairs(uint64_t x) {
    return popcount((x | x >> 1) & 0x5555555555555555);
}
\end{lstlisting}

\ttlpar{Compacting offsets and seeds.} After removing the ranking structure, it
is only left to store the partial sums of the number of keys per chunk, and the
seed used for the chunk hash function. This is the totality of the overhead
imposed by the HEM data structure with respect to constructing the function over
the whole input set at once.

Instead of storing these two numbers separately, we combine them into a single
64-bit integer. The main observation that allows us to do so is that due to the
extremely high probability of finding a good seed for each chunk, few random
bits are necessary to store it: we can just use the same sequence of seeds for
each chunk, and store the number of failed attempts before the successful
one. In our experiments this number is distributed geometrically and never
greater than $24$.
If we are storing $n$ keys, $64-\lceil \log n\rceil$ bits are available for the seed, which are more than
sufficient for any realistic $n$.

\section{Experimental results}%

We performed experiments in Java using two datasets derived from the \texttt{eu-2015} crawls
gathered by BUbiNG~\cite{BMSB} on an Intel\textregistered{} Core\texttrademark{}
i7-4770 CPU @3.40GHz (Haswell). The smaller dataset is the list of hosts
(11\,264\,052 keys, $\approx 22$\,B/key), while the larger dataset is the list
of pages (1\,070\,557\,254 keys, $\approx 80$\,B/key). The crawl data is
publicly available at the LAW website.\footnote{\texttt{http://law.di.unimi.it/}}

\begin{figure}[htb]
\centering
\includegraphics[scale=.26]{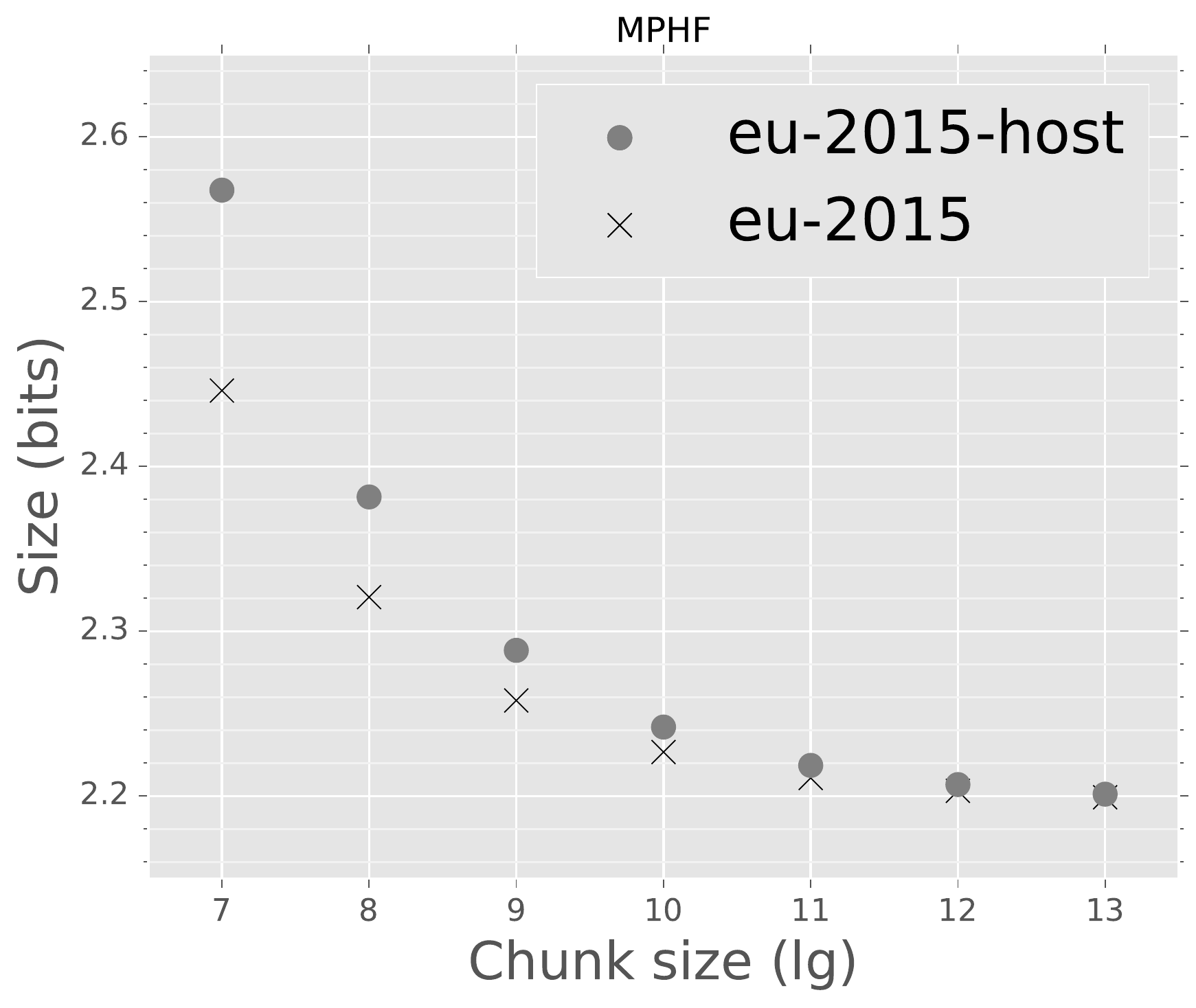}\qquad\qquad \includegraphics[scale=.26]{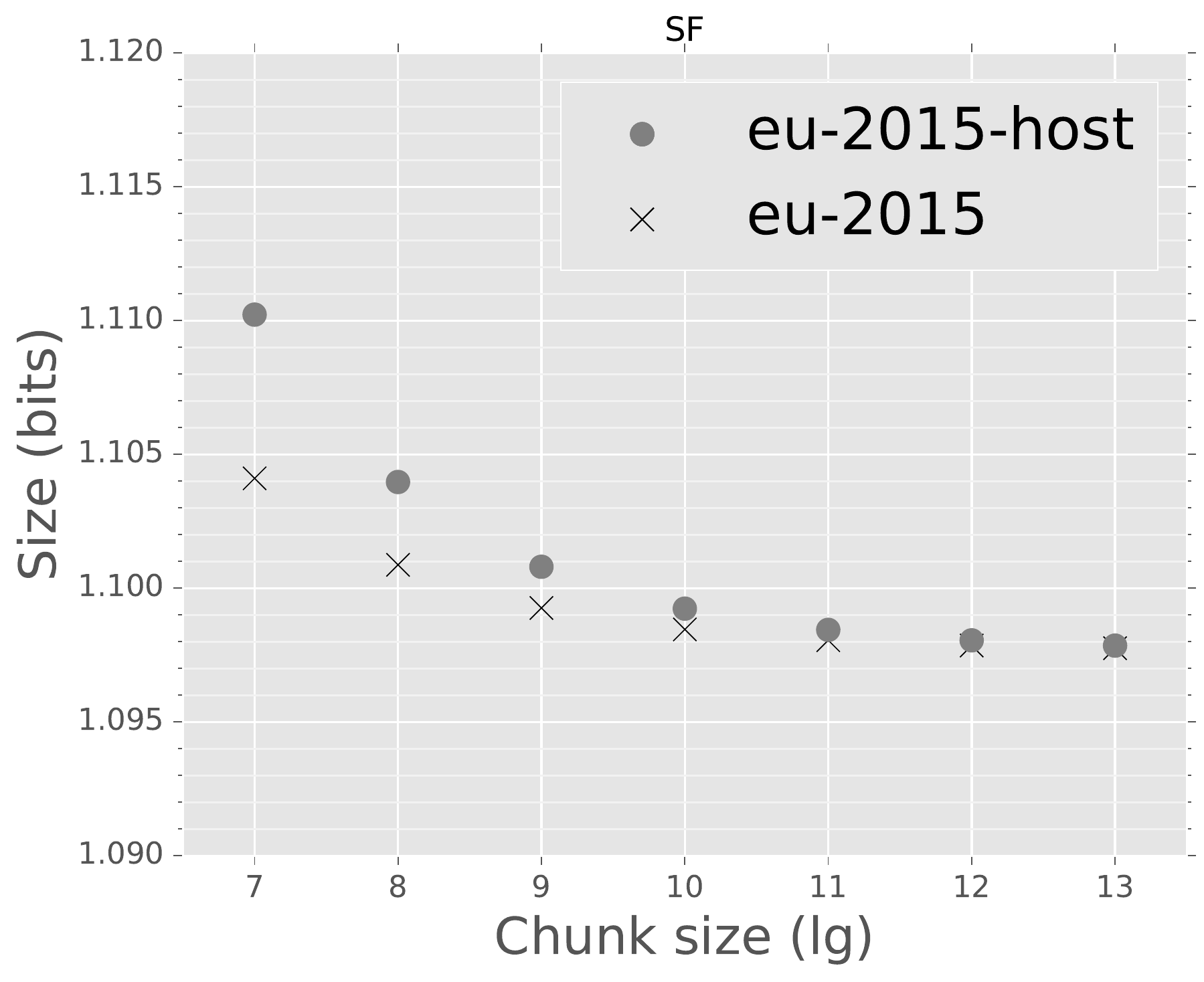}

\includegraphics[scale=.26]{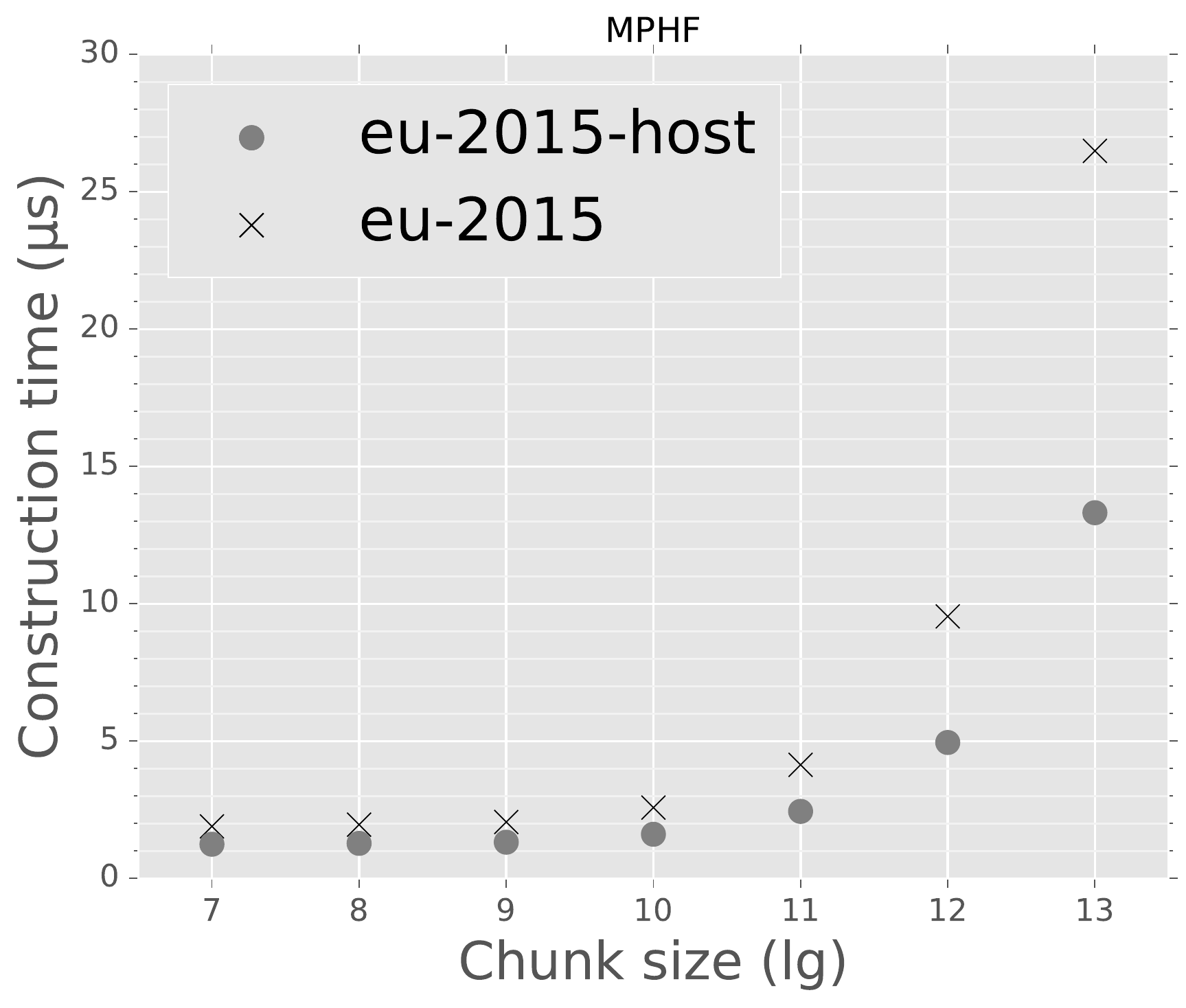}\qquad\qquad \includegraphics[scale=.26]{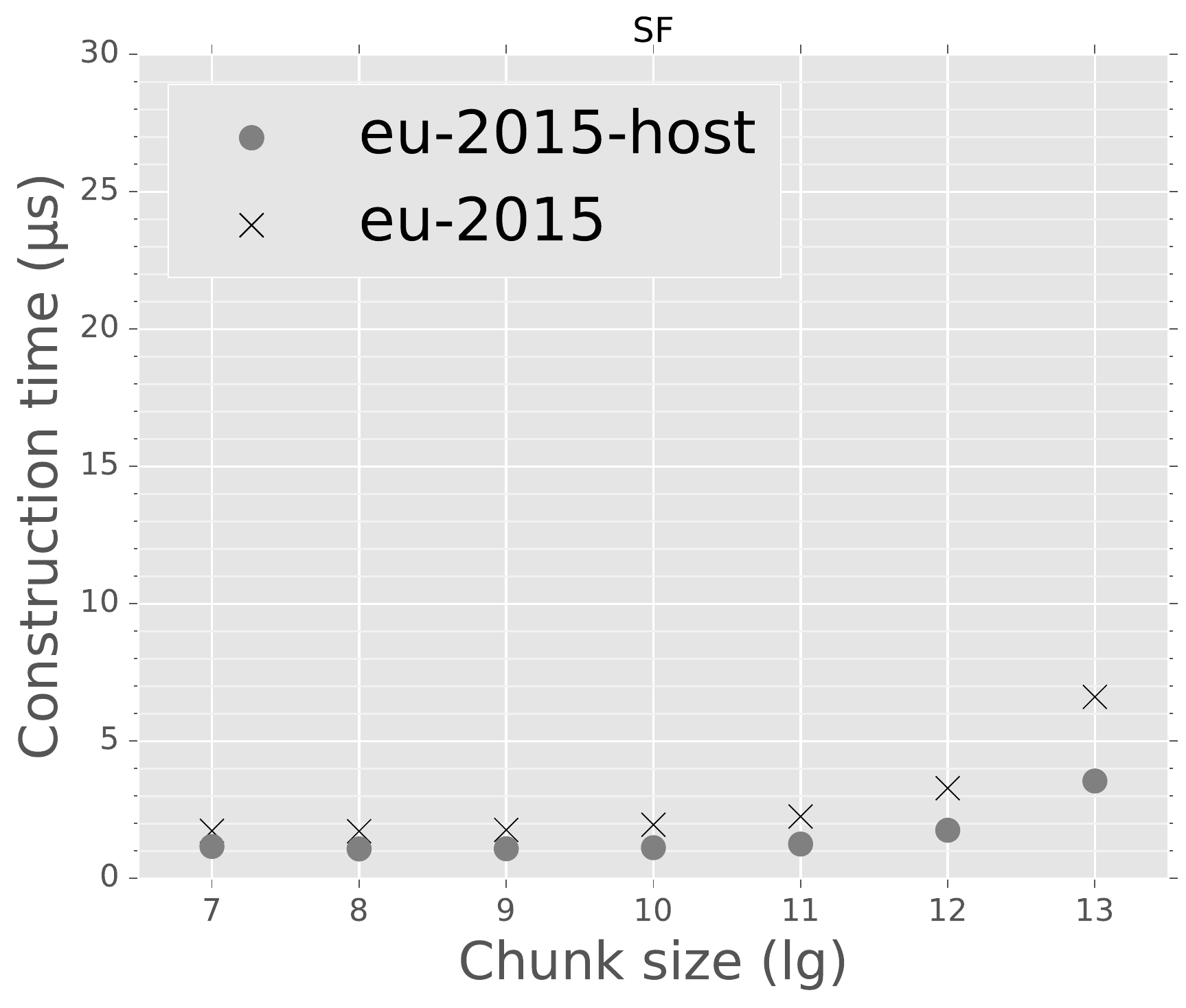} 

\includegraphics[scale=.26]{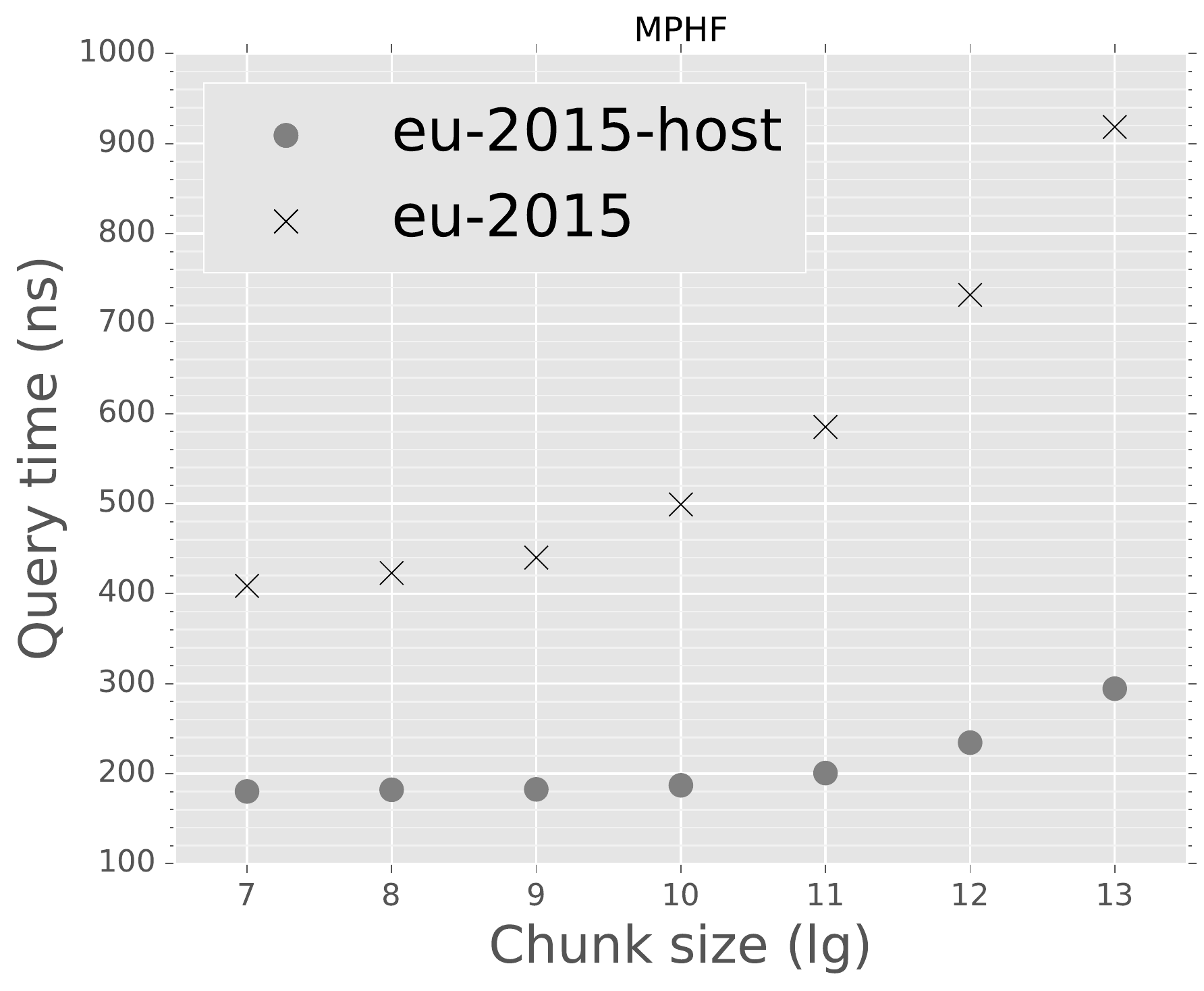}\qquad\qquad \includegraphics[scale=.26]{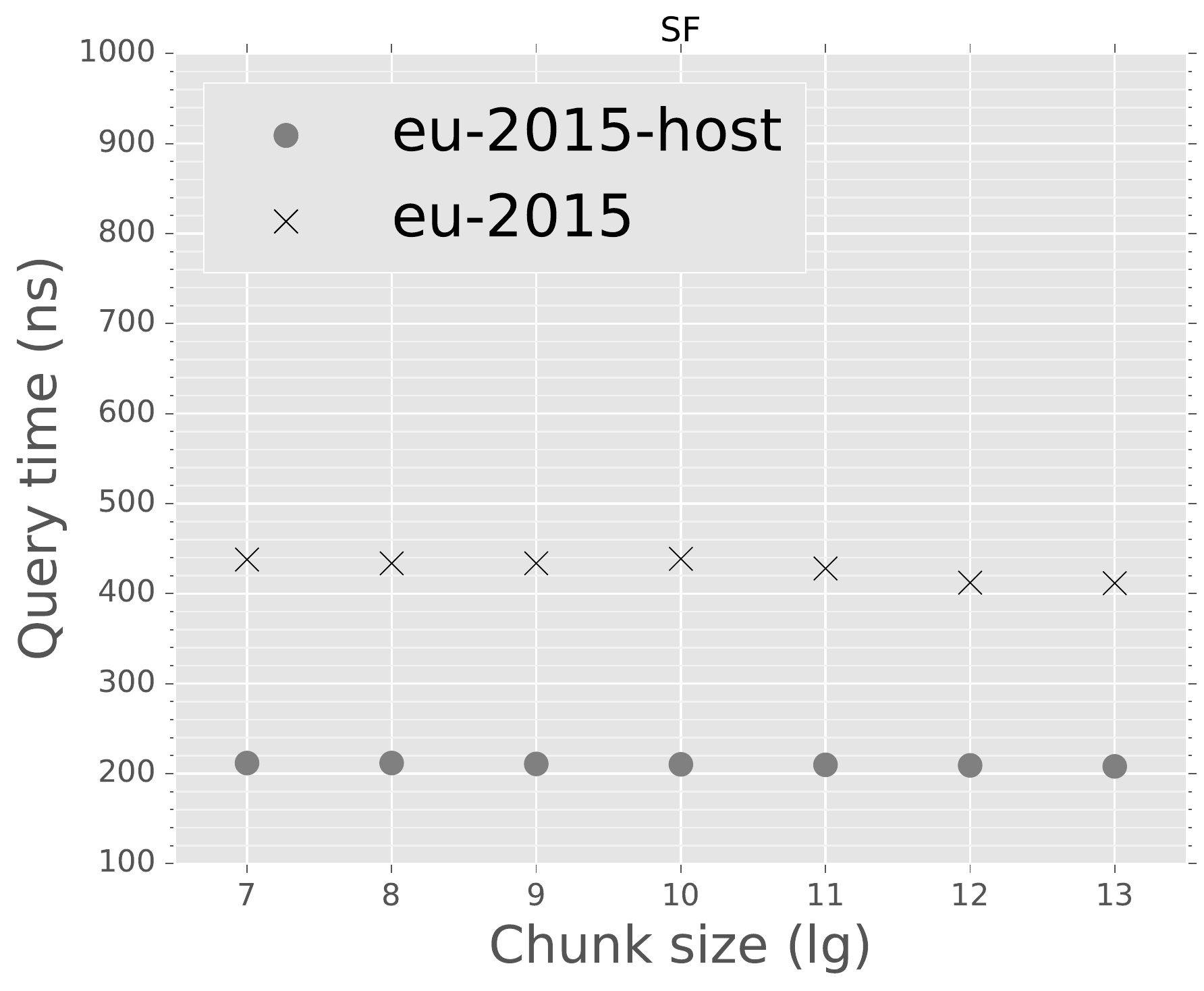}
\caption{\label{fig:measure} Size in bits per element, and construction and lookup time in microseconds for the \texttt{eu-2015}
and \texttt{eu-2015-host} datasets when $r=3$.}
\end{figure}

Besides the final performance figures (which depends on the chosen chunk size),
it is interesting to see how the measures of interest vary with the chunk size.
In Figure~\ref{fig:measure} we show how the number of bits per element,
construction time and lookup time vary with the chunk size for $r=3$. Note that
in the case of minimal perfect hash functions we show the actual number of bits
per key. In the case of general static function, we build a function mapping
each key to its ordinal position and report the number of additional bits per
key used by the algorithm.

As chunks gets larger, the number of bits per key slightly decreases (as the
impact of the offset structure is better amortized); at the same time:
\begin{itemize}
  \item construction time increases because the Gaussian elimination process is
  superlinear (very sharply after chunk size $2^{11}$);
  \item in the case of minimal perfect hash functions, larger chunks
cause the rank function to do more work linearly with the chunk size, and indeed
lookup time increases sharply in this case;
\item in the case of static functions, chunks larger than $2^{10}$ yield a
slightly improved lookup time as the offset array becomes small enough to fit
in the L3 cache.
\end{itemize}

In Table~\ref{tab:time}, we show the lookup and construction time of our ``best
choice'' chunk size, $2^{10}$, with respect to the data reported
in~\cite{ADREVTGRDS} for the same space usage (i.e., additional $1.10$\,b/key),
and to the C code for the CHD technique made available by
the authors (\texttt{http://cmph.sourceforge.net/}) when $\lambda=3$, in which
case the number of bits per key is almost identical to ours.
We remark that in the case of CHD for the larger dataset we had to use different
hardware, as the memory available (16\,GB) was not sufficient to complete the
construction, in spite of the final result being just 3\,GB.

\begin{table}[htb]
\renewcommand{\tabcolsep}{7pt}
\centering
\begin{tabular}{lccccccc}\toprule
\multicolumn{1}{c}{}        & \multicolumn{3}{c}{eu-2015-host}  &\multicolumn{3}{c}{eu-2015} &  ADR \\
\cmidrule(rl){2-4} \cmidrule(rl){5-7} \cmidrule(rl){8-8}
\multicolumn{1}{c}{}       & MPHF            & SF  &CHD          & MPHF          & SF  &CHD      &  SF\\
\midrule
Lookup (ns)  & 186 	 &210 &	408 &499  &438 &	1030 & ?\\
Construction ($\mu$s) & $1.61$ 	 &$1.12$ &$0.98$	 &$2.45$ 	 &$1.73$ & $3.53$
&$270$ 	 \\
\bottomrule
\end{tabular}
\caption{\label{tab:time} A comparison of per-key construction and evaluation
time, $r=3$. CHD is from~\cite{BBDHDC}, ADR is from~\cite{ADREVTGRDS}.}
\end{table}

\begin{table}[htb]
\renewcommand{\tabcolsep}{7pt}
\centering
\begin{tabular}{lcccc}\toprule
\multicolumn{1}{c}{}        & \multicolumn{1}{c}{eu-2015-host} &\multicolumn{1}{c}{eu-2015} &\multicolumn{1}{c}{ADR} \\
\cmidrule(rl){2-3} \cmidrule(rl){4-4}  
Lookup (ns)  & 236 & 466  & ?\\
Construction ($\mu s$) & 1.75 & 2.6 & $\approx$2000\\
\bottomrule
\end{tabular}
\caption{\label{tab:time4} Per-key construction and evaluation time of static
functions, $r=4$.} \end{table}

In the case of static function, we can build data structures about two
hundred times faster than what was previously possible~\cite{ADREVTGRDS} (the data displayed
is on a dataset with $10^7$ elements; lookup time was not reported).
To give our reader an idea of the contribution of each technique we use,
Table~\ref{tab:sub} shows the increase in construction time using any combination of
the peeling phase (which is technically not necessary---we could just solve the
system), broadword computation instead of a standard sparse system
representation, and lazy instead of standard Gaussian elimination. The combination
of our techniques brings a \emph{fifty-fold} increase in speed (our basic speed is already
fourfould that of~\cite{ADREVTGRDS}, likely because our hardware is more
recent).

\begin{table}[htb]
\renewcommand{\arraystretch}{1.3}
\centering
\caption{\label{tab:sub}Increase in construction time for $r=3$ using just
pre-peeling (P), broadword computation (B), lazy Gaussian elimination (G) or a combination.}
\begin{tabular}{c|c|c|c|c|c|c}
 BG& GP& G& BP& B& P& None\\
 \hline
 +13\%& +57\%& +98\%& +296\%& +701\%& +2218\%& +5490\%\\
\end{tabular}
\end{table}

In the case of MPHFs, we have extremely competitive lookup speed (twice that of
CHD) and much better scalability. At small size, performing the
construction entirely in main memory, as CHD does, is an advantage, but as soon
as the dataset gets large our approach scales much better. We also remark that our code is a highly abstract
Java implementation based on \emph{strategies} that turn objects into bit vectors at
runtime: any kind of object can thus be used as key. A tight C implementation able to
hash only byte arrays, such as that of CHD, would be significantly faster. Indeed,
from the data reported in~\cite{BBOCOPRH} we can estimate that it would be about twice as fast.

The gap in speed is quite stable with respect to the key size: testing 
the same structures with very short (less than 8 bytes) random keys provides of course
faster lookup, but the ratio between the lookup times remain the same. 

Finally, one must consider that CHD, at the price of a much greater construction time, can further
decrease its space usage, but just a 9\% decrease in space increases
construction time by an order of magnitude, which makes the tradeoff
unattractive for large datasets.

With respect to our previous peeling-based implementations, we increase
construction time by $\approx50$\% (SF) and $\approx100$\% (MPHF), at the same
time decreasing lookup time.

In Table~\ref{tab:time4} we report timings for the case $r=4$ (the construction
time for~\cite{ADREVTGRDS} has been extrapolated, as the authors do not provide
timings for this case).
Additional space space required now is just $\approx3$\% (as opposed to $\approx10$\% when
$r=3$). The main drawbacks are the slower construction time (as the system
becomes denser) and the slower lookup time (as more memory has to be accessed). Larger
values of $r$ are not interesting as the marginal gain in space
becomes negligible.

\section{Further applications}

Static functions are a basic building block of \emph{monotone} minimal perfect
hash functions~\cite{BBPTPMMPH2}, data structures for weak prefix
search~\cite{BBPFPSLSA}, and so on. Replacing the common MWHC
implementation of these building blocks with our improved construction will
automatically decrease the space used and the lookup time in these data structures.

We remark that an interesting application of static functions is the almost
optimal storage of \emph{static approximate dictionaries}. By encoding as a
static function the mapping from a key to a $b$-bit signature generated by a random hash function,
one can answer to the question ``$x\in X$?'' in constant time, with false
positive rate $2^{-b}$, using (when $r=4$) just $1.03nb$ bits; the lower bound
is $nb$~\cite{CFGEAMT}.

\section{Conclusions}

We have discussed new practical data structures for static functions and minimal
perfect hash functions. Both scale to billion keys, and both improve
significantly lookup speed with respect to previous constructions. In
particular, we can build static functions based on Gaussian elimination two
orders of magnitude faster than previous approaches, thanks to a combination of
broadword programming and a new, parameterless lazy approach to the solution of
sparse system. We expect that these structure will eventually replace the
venerable MWHC approach as a scalable method with high-performance lookup.

\bibliography{biblio}

\end{document}